\magnification=\magstep1
\hsize=14.6 truecm
\vsize=21.5 truecm
\baselineskip=20 pt
\null 
\def\FS{Frobenius-Schur indicator }
\def\n{\nu_p}
\def\o{\omega}
\def\e{\epsilon}

\null\vskip 3truecm
 
\centerline{\bf The \FS in Conformal Field Theory}\bigskip

\centerline{{Peter Bantay}\footnote{$^1$}{On leave from the
Institute for Theoretical Physics, E\"otv\"os University}
\footnote{$^{,2}$}{Partially supported by OTKA T016251}}\bigskip

\centerline{Mathematics Department, University of California at Santa Cruz
}\bigskip

\vskip 3truecm
 
\noindent {\it Abstract} :  An analogue of the classical \FS is
introduced in order to distinguish between real and pseudo-real self-conjugate
primary fields, and an explicit expression for this quantity is derived
from the trace of the braiding operator.

\vfill\eject

\bigskip\bigskip

The distinction between real and pseudo-real primary fields in a CFT had been
known for a long time [1]. It corresponds to the different behavior of
the holomorphic blocks of the two-point functions under braiding, and
amounts to a difference in sign. A similar pattern exists for the
self-conjugate superselection sectors of two-dimensional QFT-s and the
representations of the related Rational Hopf-algebras [2]. The aim of
this paper is to show that the above distinction is already encoded in
the standard data of the CFT under study, i.e. the fusion rules, the
quantum dimensions and the conformal weights of the primary fields.

An analogous classification exists in group representation theory,
where one distinguishes between real, complex and quaternionic
representations ( the latter are sometimes called pseudo-real )
according to the realizability of the given representation with
matrices whose elements lie in the respective (skew)-field [3]. It is
usual to introduce the so-called Frobenius-Schur indicator for
irreducible representations, which is +1 for real, 0 for complex, and
-1 for quaternionic representations, and there is a simple formula that
allows to compute it from the character of the representation. In fact,
the \FS is not directly related to the reality properties of the
representations; it is telling us whether there is an invariant
bilinear form, and if there is one, whether it is symmetric or
anti-symmetric. By a well-known theorem this may be translated into an
assertion about the reality properties.

We can try to mimic the above situation by introducing for the primary
fields of a CFT the \FS $\n$, which is +1 if the field $p$ is real,
$-1$ if it is pseudo-real, and $0$ if it is complex, i.e. different from
its charge conjugate $\bar p$.

\vfil\eject

What is truly remarkable about the \FS is that there is an explicit
formula expressing it in terms of such standard quantities as the
fusion rules, the (quantum) dimensions and the conformal weights of the
primary fields.  The explicit expression for the indicator reads
$$\nu_p=\sum_{r,s}N_{rs}^pS_{0r}S_{0s}{\o_r^2\over\o_s^2},\eqno(1)$$
where $\o_p=\exp(2\pi\imath\Delta_p)$ is the exponentiated conformal
weight of the primary field $p$.

To understand the origin of the above formula, let's consider the
braiding of $p$ with itself.  Suppose that the primary field $q$
appears in the OPE of $p$ with itself, i.e. $N_{pp}^q>0$, and let's
consider the  operator ${\cal R}_{pp}^{[q]}$ that braids the two copies
of $p$ in the intermediate channel $q$. The braiding has as its square
the monodromy which we know to be diagonal with eigenvalue $\o_q
\o_p^{-2}$, i.e.  $${\cal R}_{pp}^{[q]2}={\o_q\over \o_p^2}{\bf
1},\eqno(2)$$ from which it follows that the possible eigenvalues of
${\cal R}_{pp}^{[q]}$ are $\pm\o_q^{1/2}\o_p^{-1}$, and should we know
the trace of ${\cal R}_{pp}^{[q]}$ we would know the multiplicities of these
eigenvalues as well.  But the trace may be computed, and the result is 

$${\rm Tr}\left({\cal R}_{pp}^{[q]}\right)=
\o_p^{-1}\sum_{r,s}N_{rs}^pS_{qr}S_{0s}{\o_s^2\over\o_r^2}\eqno(3)$$

We are now ready to express the multiplicity of the eigenvalue
$\pm\o_q^{1/2}\o_p^{-1}$ of ${\cal R}_{pp}^{[q]}$ as

$${1\over 2}\left(N_{pp}^q\pm\o_q^{-1/2}
\sum_{r,s}N_{rs}^pS_{qr}S_{0s}{\o_s^2\over\o_r^2}\right)\eqno(4)$$
Note that by the above reasoning the value of the expression 

$${\cal Z}(p,q)=
\o_q^{-1/2}\sum_{r,s}N_{rs}^pS_{qr}S_{0s}{\o_s^2\over\o_r^2}\eqno(5)$$  
should be an integer of the same parity as $N_{pp}^q$ and whose absolute
value is not greater than $N_{pp}^q$. We remark that these integrality
conditions could prove very effective in the classification program,
because they rule out many possibilities for  prospective fusion rule
algebras that pass all other tests.

Returning to the \FS, we have to consider the special case $q=0$ in the
above reasoning, because we are interested in the behavior of the
two-point function.  In this case $N_{pp}^0=\delta_{p,\bar p}$, and it
follows from the remarks following $(5)$ that ${\cal Z}(p,0)=0$ if
$p\ne \bar p$, and ${\cal Z}(p,0)=\pm 1$ if $p=\bar p$ depending on the
eigenvalue of the braiding, i.e. the symmetry ( or anti-symmetry ) of
the two-point block.  But this is exactly the three-fold
classification that led us to define the indicator, proving our assertion
that $\n={\cal Z}(p,0)$.

The \FS is especially simple to compute for simple currents, i.e.
primary fields whose (quantum) dimension is 1. It turns out that in
this case 
$$\nu_\e=\o_\e^2 \delta_{\e,\bar \e }\eqno(6)$$ 
That is, the conformal weight of a real simple current is an integer or
half-integer, while that of a pseudo-real one is $\pm{1\over 4}$ modulo
integers. Actually, Eq.(6) is but a special case of the more general
relation
$$\nu_{\e p}=\o_\e^2 \nu_p,\eqno(7)$$
valid for any self-conjugate simple current $\e$.

\vfil\eject

\underbar{References} :

[1] Alvarez-Gaume, Gomez and Sierra : Nucl. Phys. {\bf B319}, (1989),
155.
 
   Moore and Seiberg :  Phys. Lett {\bf B212} (1988), 451.

[2] Fredenhagen, Rehren and Schroer : Rev. Math. Phys. 113, spec. issue
(1992).
 
   Fuchs, Ganchev and Vecsernyes : International Journal of Modern
Physics {\bf A10}, (1995), 3431.

[3] Isaacs, I. M. : Character Theory of Finite Groups, Academic Press, 1976.
   
 Kirillov, A. A. : Elements of the Theory of Representations, Springer, 1976.

\end